\documentclass[3pt,onecolumn]{article}
\usepackage{graphicx}

\newcommand{\ben}{\begin{enumerate}}
\newcommand{\een}{\end{enumerate}}

\newcommand{\be}{\begin{equation}}
\newcommand{\ee}{\end{equation}}
\newcommand{\bea}{\begin{eqnarray}}
\newcommand{\eea}{\end{eqnarray}}

\begin{document}
%\renewcommand{\ee}{\end{equation}}
%\begin{flushright}
%Corrected version, April 2010\\
%Not for further distribution\\
%\end{flushright}
\vspace{0.1cm}
\thispagestyle{empty}

%\newpage
\begin{center}
%{\Large\bf }\\[13mm]
{\Large\bf 
The geometrical origin of the strain-twist coupling in double helices}\\[13mm]
%xxxxxxx12345678902234567890323456789042345678905234567890623456789072345678908234567890923456
{\rm Kasper Olsen{\footnote{kasper.olsen@fysik.dtu.dk}} and Jakob Bohr\footnote{jakob.bohr@fysik.dtu.dk}}\\[2.5mm]
{\it Department of Physics,\\ Technical University of Denmark}\\
{\it Building 307 Fysikvej, DK-2800 Lyngby, Denmark}\\[6mm]
%\end{center}
%{\sc Abstract}\\
\end{center}

%\begin{article}

\begin{abstract}
The geometrical coupling between strain and twist in double helices is investigated. Overwinding, where strain leads to further winding, is shown to be a universal property for helices, which are stretched along their longitudinal axis when the initial pitch angle is below the zero-twist angle (39.4$^\circ$). Unwinding occurs at larger pitch angles. The zero-twist angle is the unique pitch angle at the point between overwinding and unwinding, and it is independent of the mechanical properties of the double helix. 
This suggests the existence of zero-twist structures, i.e. structures that display neither overwinding, nor unwinding under strain. Estimates of the overwinding of DNA, chromatin, and RNA are given.
\end{abstract}\vspace{0cm}

%\keywords{Strain | Twist | Mechanics | Helical symmetry | Overwinding | Double helix }

%\end{article}
%\newpage
\section{Introduction}

If one pulls a double helix structure by the end, one might think that it would unwind by the applied tension. In this paper we show why this is not always the case: A helix can unwind, overwind, or it can stay at its current twist (which we denote a zero-twist (ZT) structure). Overwinding is contrary to unwinding; unwinding is the de-twisting of the helices obtained by stretching the material. For the zero-twist structure there is no coupling from strain to twist.
The existence of a twist-stretch coupling is a well-known phenomenon for helical steel wires \cite{utting1987} where it leads to unwinding, and design efforts go into designing rotation resistant wire rope when desired \cite{pellow1982,waterhouse2003}.

The geometrical investigation presented below is based on the study of packed double helices modeled as two flexible tubes with hard walls. To be packed is defined by the constraint of the two tubes being in contact. 
Does this mean that the helices are stretched? No, generally not, stretching is one way 
to secure that a packed helix is obtained, however, for helices on the molecular size 
favorable molecular interactions can also make it more preferable to be packed than not. 
A detailed analysis of packed helices and their volume fractions showed that the helices with the highest volume fractions are noticeably similar to the molecular structure of DNA \cite{olsen2009}; this suggests that close-packing is at work as a structure forming principle. For the description of compact strings and tube models, the importance of one kind of optimum shape has been discussed by Gonzalez and Maddocks \cite{gonzalez1999} and Maritan et al. \cite{maritan2000}, and one related suggestion for the best packing of proteins and DNA has been considered by Stasiak and Maddocks \cite{stasiak2000}. A detailed analysis of the geometry of $n$-plies, and of their self-contacts, has been given by Neukirch and van der Heijden \cite{neukirch2002}.

\section{Model}

The close-packed (CP) structure with an optimized volume fraction has a pitch angle of $32.5^\circ$ \cite{olsen2009}: This structure that has a central channel is shown in Figure 1a. Under a pull, the pitch angle is increased and the diameter of the central channel gets smaller, and eventually, the inner channel disappears at a pitch angle of $45 ^\circ$. Whether a helix overwinds or unwinds is then determined from the balance between the gain in length from the reduction in the helical radius versus untwisting. The crossing point -- which we denote as the {\it zero-twist} angle -- is at $39.4^\circ$ (Figure 1b) and is smaller than the $45^\circ$, where the helical radius becomes equal to the diameter of the tubes, and
is maintained for all pitch angles above $45 ^\circ$. 
The $45^\circ$ motif, here denoted the tightly packed (TP) double helix, is shown in Figure 1c.

\begin{figure}[h]
\begin{center}
\includegraphics[width=8.5cm]{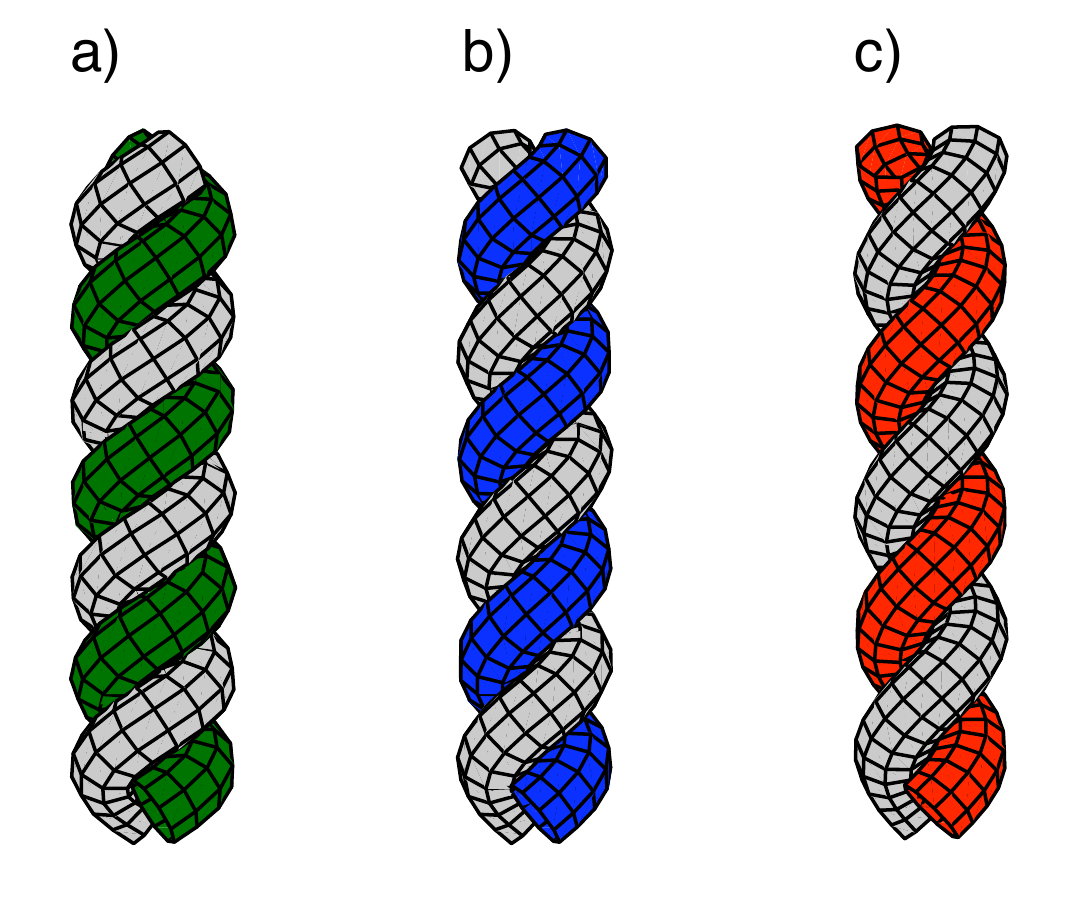}
\caption{{\it Different geometries of a double helix of tubes of fixed diameter $D$. {\bf a)} Close-packed (CP) structure of pitch angle $32.5^\circ$ measured from horizontal. {\bf b)} Zero-twist (ZT) structure with a pitch angle of $39.4^\circ$. It is at the point between overwinding/unwinding. {\bf c)} Tightly-packed (TP) structure of pitch angle $45^\circ$. Overwinding from stretching takes place from the a) to the b) confirmations; unwinding from b) to c).}}
\end{center}
\end{figure}

Geometrically, the double helix is given by two tubes of diameter $D$, whose centerline defines two helices with simple parametric equations. A helix is a curve of constant curvature, $\kappa$, and torsion, $\tau$, %\cite{stoker1969}; 
and it can be specified by two parameters, for example $a$ and $H$, where $a$ is the helix radius (the radius of the cylinder hosting the helical lines) and $H$ the helical pitch (the raise of the helix for each $2\pi$ rotation). The tangent to each of the helical curves is at an angle $v_\bot$ ({\it the pitch angle}) with the horizontal axis, and it is determined by $\tan v_\bot = h/a$, where $h=H/2\pi$ is the reduced pitch.
We say that the double helix is packed when the shortest distance between the centerline of one helical tube to the next one equals the diameter $D$ of the tubes, i.e. the double helix is packed when the tubes are in contact. The volume fraction can be calculated using, as a reference volume, an enclosing cylinder of height $H=2\pi h$ and volume $V_E=2\pi^2 h(a+D/2)^2$, and comparing it to the combined volume occupied by the two circumscribed helical tubes, $V_H = \pi^2 h D^2 / \sin v_\bot$. The volume fraction is the ratio of the two volumes, i.e. $f_V = V_H/V_E$, which reads
\be
\label{}
f_V 
= 2 (1+(\frac{a}{h})^2)^{1/2} \cdot (\frac{2a}{D}+1)^{-2}
\ee

\noindent With this choice of reference volume the packing fraction depends only on the shape of the double helix structure, which can be described by one parameter, e.g. the pitch angle, $v_\bot$.
The maximum of $f_V$ defines the close-packed (CP) helix. For the double helix this maximum is at $v_\bot^* = 32.5^\circ$, where $f_V^*=0.796$ \cite{olsen2009}. 
For the CP structure, the channel radius is about 17~\% of $a$ \cite{olsen2009}.
Generally, the radius of the central channel, which is given by $R_i=a-D/2$, is a decreasing function of $v_\bot$; this can be seen from Figure 2 which shows $2a/D$ depending on the pitch angle.
For $v_\bot \geq 45^\circ$ there is no central channel as $2a/D=1$, see Figure 2.

\begin{figure}[h]
\begin{center}
\includegraphics[width=9.5cm]{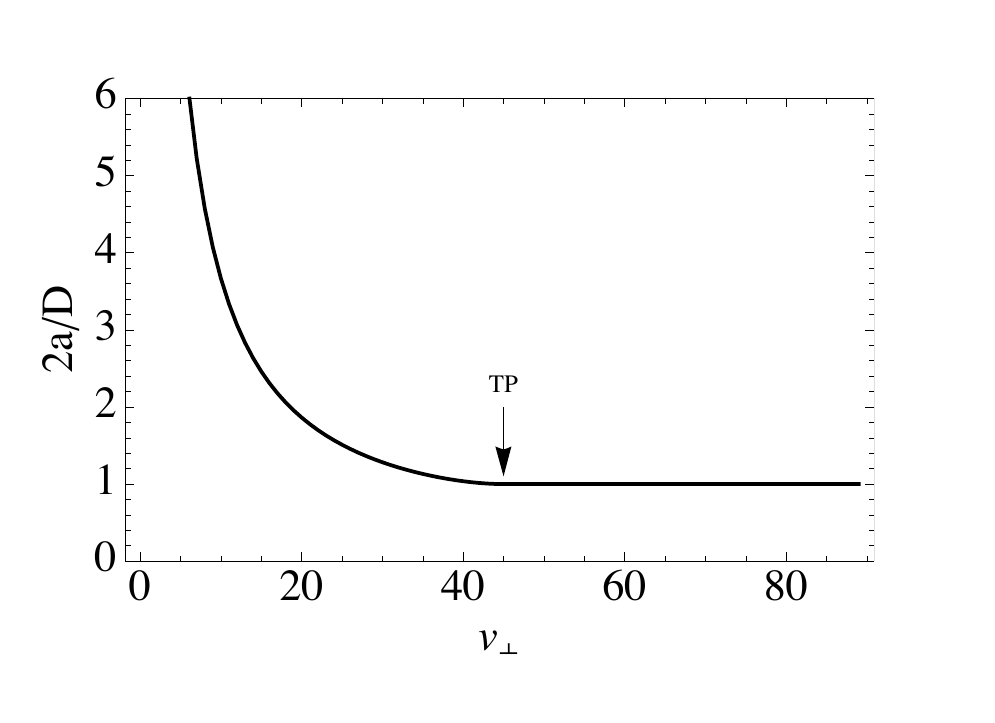}
\caption{{\it Graph showing the ratio $2a/D$ as a function of pitch angle, $v_\bot$ [deg.], where $a$ is the helix radius and $D$ the diameter of the helical tubes. The tightly packed double helix has a pitch angle of $v_{TP}=45^\circ$; it is the helix with the smallest pitch angle obeying the criterion that $2a=D$.}}
\end{center}
\end{figure}

\section{Results}

Consider a long straight segment of a double helix consisting of two long molecular strands each of length $L_M$. The length of the double helix is $H_M=L_M\sin v_\bot$ and the total twist is $\Theta_M=L_M\cos v_\bot/a$. In Figure 3 the 
dimensionless ratio $D\theta_M/2L_M$ is shown as a function of the pitch angle. One can see that for $v_\bot<v_{ZT}$ there is overwinding while for $v_\bot > v_{ZT}$ there will be unwinding. We find numerically that $v_{ZT}=39.4^\circ$.

\begin{figure}[h]
\begin{center}
\includegraphics[width=9.5cm]{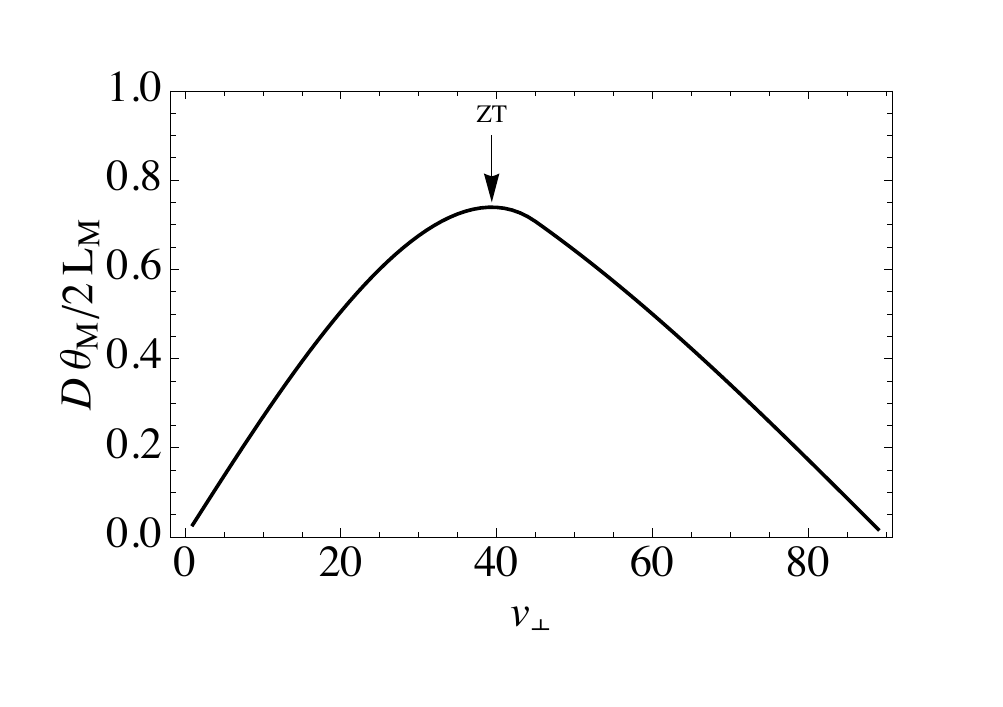}
\caption{{\it The total twist, $\theta_M$, for a long segment of the double helix; the dimensionless quantity 
$D\theta_M /2 L_M$ is shown as a function of the pitch angle, $v_\bot$ [deg.]. The maximum value is obtained for the pitch angle $v_{ZT}=39.4^\circ$ and mark the transition from overwinding to unwinding. At the ZT structure there is zero coupling between twist and strain.}}
\end{center}
\end{figure}

We can determine the amount of overwinding and unwinding in the following way. If a long double helical segment is stretched a bit, the pitch angle, $v_\bot$, will change by a small amount
$dv_\bot$, and hence $H_M$ changes by
\begin{equation}
dH_M=L_M\cos v_\bot d v_\bot
\end{equation}
and $\Theta_M$ by
\begin{equation}
d\Theta_M=- L_M\frac{\sin v_\bot}{a}dv_\bot- \frac{L_M}{a^2}\cos v_\bot\frac{da}{dv_\bot}dv_\bot
\end{equation}
so that
\begin{equation}
\label{ow}
\frac{d\Theta_M}{dH_M}=-\frac{1}{a}\tan v_\bot -\frac{1}{a^2}\frac{da}{dv_\bot}
\end{equation}
If this derivative is positive, then the helix will overwind, and if it is negative, it will unwind.
The derivative in Eq. (\ref{ow}) has dimension of inverse length. From a geometrical viewpoint it is more natural to look at the dimensionless function of $v_\bot$, obtained by multiplying with the common radius of the tubes, $(D/2)$, namely:
\begin{equation}
\label{ow2}
\frac{D}{2}\frac{d\Theta_M}{dH_M}=-\frac{D}{2a}\tan v_\bot +\frac{d}{dv_\bot}\left( \frac{D}{2a}\right)
\end{equation}
This equation can be given a simple interpretation. The first term is negative and determines the amount of unwind, while the second term is positive and determines the amount of overwind.
The graph of this derivative, that dictates the coupling between strain and twist, is depicted in Figure 4. Notice that the CP double helix will always overwind since $d\Theta_M/dH_M >0$. At the close-packed structure, the overwind is $(D/2) d\Theta_M/dH_M = 0.665$. The extension is therefore universally determined just by giving the diameter, $D$, of the tubes making up any close-packed double helix. At the zero-twist structure, $v_{ZT}=39.4^\circ$, there is neither overwinding, nor unwinding.
For larger pitch angles the overwind, $(D/2)d\Theta_M/dH_M$, is negative and the double helix will unwind under strain. 
It is therefore crucial, that the pitch angle is below that of the zero-twist ($39.4^\circ$) for overwinding to be observed, but it also indicates that elastic properties of the material are not essential to understanding the phenomenon.

\begin{figure}[h]
\begin{center}
\includegraphics[width=10.cm]{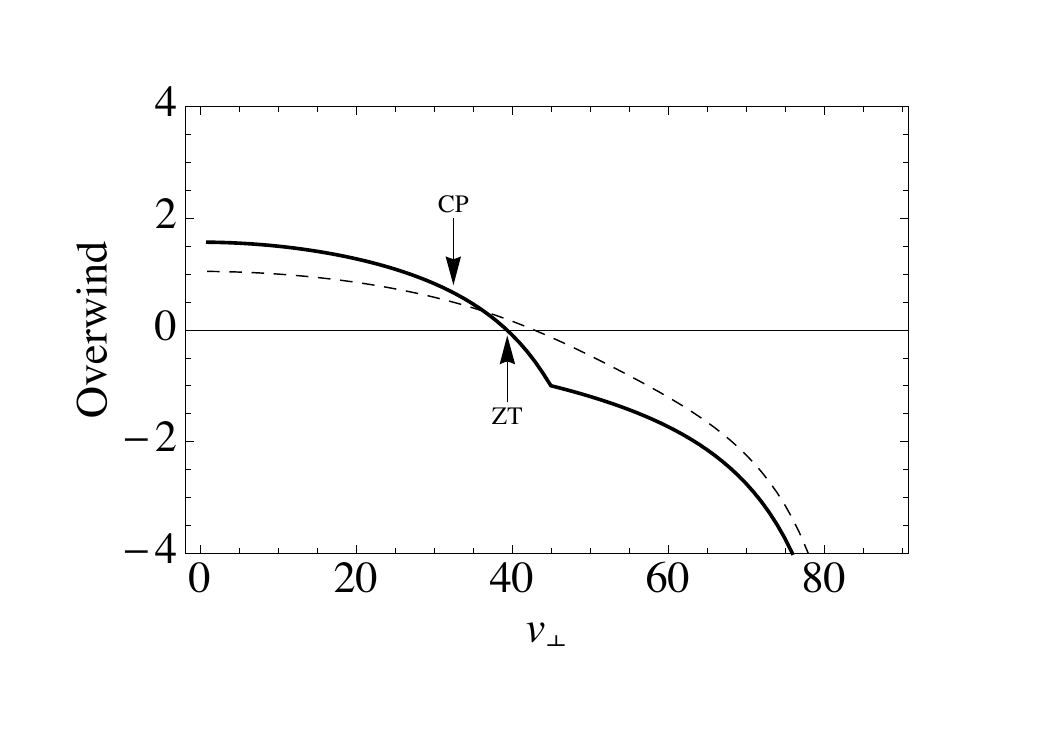}
\caption{{\it Graph showing the calculated overwind of double helices (solid line), i.e. Eq. (\ref{ow2}) as a function of $v_\bot$ [deg.]. A positive overwind means that the double helix will exhibit overwinding, while a negative overwind means that the double helix will exhibit unwinding. The zero-twist structure (ZT) is indicated with an arrow at $v_{ZT}=39.4^\circ$, the close-packed structure (CP) is indicated by an arrow at $v_{CP} = 32.5^\circ$. The first derivative is discontinuous at $v_{TP}=45^\circ$ where the helix radius can not get smaller. The dashed line is the overwind for a triple helix, which has a zero-twist angle of $42.8^\circ$.}}
\end{center}
\end{figure}

\section{Discussion}

%As a simple test of the validity of our derived strain-twist coupling, we double twisted a household rope with diameter $D$ of about $11$ mm by twenty full rotations over a length of 80 cm. We then observed the extension of the rope while it was being unwinded one full $2\pi$ rotation. The extension was about 29 mm and thus in agreement with the results of Figure 2, i.e. $\sim \pi (1.1)^{-1} \times 11$~mm = 31 mm. 

In the following we discuss some molecular examples. 
The phenomenon of overwinding in DNA was first observed in 2006, see Lionnet et al. \cite{lionnet2006} and Gore et al. \cite{gore2006} using magnetic tweezers to control the wringing \cite{lionnet2006} and optical tweezers to control the pulling \cite{gore2006}: For small deformations, DNA overwinds when stretched, i.e. it rotates counter to unwinding. During overwinding the extension of a long chain of DNA-B has been reported to be $0.42 \pm 0.2$~nm per $2\pi$ rotation \cite{lionnet2006} and $0.5$~nm per $2\pi$ rotation \cite{gore2006}. Very recently, it has been suggested that in the absence of tension DNA is an order of magnitude softer  \cite{fenn2008}.

Using the above mathematical solution for the double helical structure of DNA we find the change of length $\Delta H$ to be
determined by
\begin{equation}
\Delta H = \frac{dH_M}{d\Theta_M} \Delta \Theta
\end{equation} 

\noindent
The diameter of the molecular tubes that make up the DNA helix is $D=1.15$ nm, which is given from our previous analysis of the close-packed structures \cite{olsen2009}. We then estimate $\Delta H$ per full $2\pi$ turn to be $\pi (0.665)^{-1} \times 1.15$~nm~$=5.4$~nm, see Figure 4. Our result seems to support the findings of ref. \cite{fenn2008}.
The geometrical restriction imposed by base pairing and its influence on $d\Theta_M/dH_M$ has not been taken into account. 
The numerical analysis has been performed for the symmetrical double helix where the close-packed structure has a pitch angle of 32.5$^\circ$. The asymmetrical DNA-B has a close-packed pitch angle of 38.3$^\circ$ and, as one can show, a zero-twist angle of 41.8$^\circ$. Theoretical work on understanding the overwinding of DNA  has focused on constructing elastic models which show a negative twist-stretch coupling \cite{sheinin2009} and on incorporating stochastic effects \cite{bernido2007}. One elastic model was considered by Gore et al. \cite{gore2006}, and consists of a rod with a stiff helical ÔwireÕ (analogous to the sugar-phosphate backbone) attached to its surface. As this system is stretched, the inner rod decreases in diameter and the helix will overwind. Smith and Healey has argued that a linear material law is inadequate for the description and suggest a non-linear elastic rod  \cite{smith2008}. 

For chromatin, the above results can be related to recent experiments in twisting chromatin fibers, see e.g. \cite{bancaud2006,celedon2009}. For a close-packed 30 nm chromatin fiber, in the so-called two-start geometry, we estimate a tube diameter of $30/(2a/D +1)$~nm= $30/(1.2+1)=13.6$~nm, where $2a/D$ is determined from Figure 2. For the close-packed 30 nm chromatin structure we then estimate $\Delta H$ per full $2\pi$ turn to be $\pi (0.665)^{-1} \times 13.6$~nm~$=64$~nm.  It is interesting to note that the numbers reported in ref. \cite{celedon2009} are measurements of $\Delta H$  for {\it Xenopus} chromatin per turn at a pulling force of 0.3 pN. Using the depicted data in ref. \cite{celedon2009} we have estimated an average extension of $\sim 60 \pm 40$~nm per turn.
Here, we have assumed the two-start helix to behave like a tubular packed double helix -- that is a view which ignores the intricate details of the structure, details which are discussed for example by Barbi et al. \cite{victor2004}, where elaborate mechanical models are described, including one which maintain its twist while being stretched.

We have presented a simple geometrical explanation for overwinding of helices -- an effect which has been observed before for the double helix of DNA and for chromatin, and which is contrary to usual unwinding. Our model of unwinding and overwinding can be applied to any symmetric double helix which is packed in the sense that the two helices touch each other, i.e. remain at the distance $D$ from each other. 
Packed double helical structures will show an overwinding behavior  similar to those already observed, as long as their initial pitch angle is sufficiently small. Perhaps, the analysis will be relevant for other helical structures such as nanofabricated quartz cylinders \cite{deufel2007}, fabricated twisted polymer nanofibers \cite{gu2007}, and for the beautiful double helical structures formed from helical carbon nanotubes \cite{liu2003}. Further, the phenomenon may be important for some aspects of the working of molecular motors during gene expression and regulation \cite{michaelis2009}. The analysis presented in this paper is straightforwardly applicable to RNA double helices \cite{baeyens1995}, which we therefore predict will show overwinding. Using a value of $26$ \AA \cite{varshavsky2006} for the molecular diameter of the double helix, we estimate an overwinding of 5.6 nm. Necturus chromatin fibers \cite{williams1986} are known to pack as a double helix with a pitch angle of $v_\bot = 32 \pm 3^\circ$ a value suggestive of being close-packed. Thus it follows that these chromatin double helices will overwind as well (other chromatin fibers with a different linker length would not necessarily overwind). Such predictions for overwinding and unwinding can nowadays be studied on single biomolecules using magnetic traps \cite{meglio2009}. 
Furthermore, the derived geometrical expressions for overwinding are straightforwardly extended to helices with more than two strands. In Figure 4 we have shown the solution for a triple helix (dashed line) which has a zero-twist angle of 
$42.8^\circ$.

Maybe one will even find examples, where Nature has build zero-twist structures, i.e. structures that display neither overwinding, nor unwinding. Chromatin with an appropriate linker length, 
and collagen are possible candidates for structures with such properties. 

\subsection*{Acknowledgements}
We would like to thank Jean-Marc Victor for helpful comments on a first version of the manuscript.

\newpage 
%\begin{large}
%\noindent {\bf References}
%\end{large}
%\section{References}

\end{document}